\documentclass[a4paper,11pt]{article}
\usepackage{pos}
\usepackage{subfig}

\begin{document}

\title{Non-perturbative bounds for $B \to D^{(*)}\ell\nu_{\ell}$ decays and phenomenological applications}
\ShortTitle{Non-perturbative bounds for $B \to D^{(*)}\ell\nu_{\ell}$ decays}

\author[a]{G. Martinelli}
\author[b]{S. Simula}
\author*[c,d]{L. Vittorio}

\affiliation[a]{Physics Department and INFN Sezione di Roma La Sapienza,\\
  Piazzale Aldo Moro 5, 00185 Roma, Italy}
  
\affiliation[b]{Istituto Nazionale di Fisica Nucleare, Sezione di Roma Tre,\\
Via della Vasca Navale 84, I-00146 Rome, Italy}

\affiliation[c]{Scuola Normale Superiore,\\
Piazza dei Cavalieri 7, 56126 Pisa, Italy}

\affiliation[d]{Istituto Nazionale di Fisica Nucleare, Sezione di Pisa,\\
Largo Bruno Pontecorvo 3, I-56127 Pisa, Italy}

\emailAdd{guido.martinelli@roma1.infn.it}
\emailAdd{silvano.simula@roma3.infn.it}
\emailAdd{ludovico.vittorio@sns.it}

\abstract{We show how to extract the Cabibbo-Kobayashi-Maskawa (CKM) matrix element $\vert V_{cb} \vert$ from exclusive semileptonic $B \to D^{(*)}$ decays by using the Dispersive Matrix (DM) method. It is a new approach which allows to determine in a full non-perturbative way the hadronic form factors (FFs) in the whole kinematical range, without making any assumption on their dependence on the momentum transfer. We investigate also the issue of Lepton Flavor Universality (LFU) by computing a pure theoretical estimate of the ratio $R(D^{(*)})$. Our approach is applied to the preliminary LQCD computations of the FFs, published by the FNAL/MILC \cite{Aviles-Casco:2019zop} and the JLQCD \cite{Kaneko:2019vkx} Collaborations, for the $B \to D^*$ decays and to the final ones, computed by FNAL/MILC \cite{Bailey_2015}, for the $B \to D$ transitions . Since the FNAL/MILC Collaborations have recently published the final results of their LQCD computations of the FFs \cite{FermilabLattice:2021cdg} for the $B \to D^*$ case, we present also the results of our procedure after its application on these data. We find $ \vert V_{cb} \vert = (41.0 \pm 1.2) \cdot 10^{-3}$ and $\vert V_{cb} \vert = (41.3 \pm 1.7) \cdot 10^{-3}$ from $B \to D$ and $B \to D^*$ decays, respectively. These estimates are consistent within $1\sigma$ with the most recent inclusive determination $\vert V_{cb}\vert_{incl} = (42.16 \pm 0.50) \cdot 10^{-3}$ \cite{Bordone:2021oof}. Furthermore, we obtain $R(D) = 0.289(8)$ and $R(D^*) = 0.269(8)$, which are both compatible with the latest experimental averages \cite{HFLAV} at the $\sim$1.6$\sigma$ level.}

\FullConference{%
 The 38th International Symposium on Lattice Field Theory, LATTICE2021
  26th-30th July, 2021
  Zoom/Gather@Massachusetts Institute of Technology
}


\maketitle

\section{State-of-the-art of semileptonic $B \to D^{(*)}$ decays}

\begin{figure}[h!]
 \centering
\includegraphics[width=0.6\textwidth]{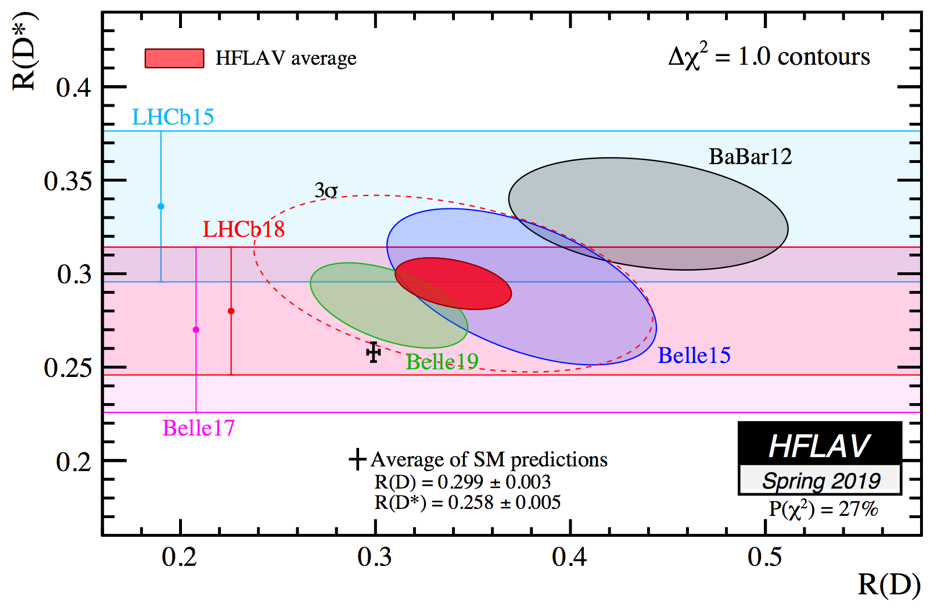}
 \centering
\caption{\textit{Correlation plot of $R(D)$-$R(D^*)$ given by the HFLAV Collaboration \cite{HFLAV} in the Spring 2019. The red area represents the world average of all measurements, while the black cross is the average of the SM predictions. The plot shows a $\sim3.1\sigma$ tension between theory and experiments.}
\hspace*{\fill} \small}
\label{RDHFLAV}
\end{figure}

Semileptonic $B \to D^{(*)} \ell \nu$ decays are one of the most challenging processes in the phenomenology of flavor physics. They are affected by two unsolved problems. On the one hand, there is the $\vert V_{cb} \vert$ \emph{puzzle}, $i.e.$ the tension between the inclusive and the exclusive determinations of the CKM matrix element $\vert V_{cb}\vert$. According to the FLAG Collaboration \cite{Aoki:2019cca}, there is a $\sim2.6\sigma$ discrepancy between the exclusive estimates (that depend on the FFs parametrization)
\begin{equation}
\vert V_{cb} \vert_{BGL} \times 10^3 = 39.08(91),\,\,\,\,\,\,\,\,\,\,\vert V_{cb} \vert_{CLN} \times 10^3 = 39.41(60),
\end{equation}
and the inclusive one
\begin{equation}
\label{inclFLAG}
\vert V_{cb} \vert_{incl} \times 10^3 = 42.00(65).
\end{equation}
Note that a new more precise estimate of the inclusive value has recently appeared \cite{Bordone:2021oof}, namely $\vert V_{cb} \vert=42.16(50)$, and it is compatible with the value quoted by FLAG in Eq.\,(\ref{inclFLAG}). On the other hand, a tension exists between the theoretical expectation value and the measurements of $R(D^{(*)})$, defined as the $\tau/\mu$ ratios of the branching fractions of $B \to D^{(*)} \ell \nu$ decays. The HFLAV Collaboration \cite{HFLAV} has produced the plot in Fig.\,\ref{RDHFLAV}, where the red area represents the average of the most recent experimental measurements. There is a $\sim3.1\sigma$ tension between that area and the Standard Model (SM) predictions, represented by a black cross in Fig.\,\ref{RDHFLAV}.

\section{The DM description of the FFs entering semileptonic $B \to D^{*}$ decays}

Let us focus our attention on semileptonic $B \to D^{*}$ decays, which are more challenging ones given the vector nature of the $D^*$ meson. The theoretical expression of the differential decay width in the massless lepton limit reads
\begin{equation}
\begin{aligned}
\label{finaldiff333BDst}
&\frac{d\Gamma(B \rightarrow D^{*}(\rightarrow D\pi) \ell \nu)}{dw d\cos \theta_{\ell} d\cos \theta_v d\chi} = \frac{G_F^2 \vert V_{cb} \vert^ 2 \eta_{EW}^2}{4(4\pi)^4} 3 m_B m_{D^*}^2 (1-2 w\,r+r^2) \sqrt{w^2-1}\, B(D^{*} \rightarrow D\pi)\\
& \hskip 3.98truecm \times  \{ (1-\cos \theta_{\ell} )^2 \sin^2 \theta_v \vert H_{+} \vert^2 + (1+\cos \theta_{\ell} )^2\sin^2 \theta_v \vert H_{-} \vert^2+ 4 \sin^2 \theta_{\ell}\cos^2 \theta_v\vert H_{0} \vert^2\\
&\hskip 3.98truecm - 2 \sin^2 \theta_{\ell}\sin^2 \theta_v \cos 2\chi  H_{+}  H_{-} - 4 \sin \theta_{\ell} (1-\cos \theta_{\ell} ) \sin\theta_v\cos\theta_v\cos\chi H_{+}  H_{0}\\
&\hskip 3.98truecm +   4 \sin \theta_{\ell} (1+\cos \theta_{\ell} ) \sin\theta_v\cos\theta_v\cos\chi H_{-}  H_{0}\},
\end{aligned}
\end{equation}
where $w$ is the recoil variable and we have introduced the helicity amplitudes 
\begin{equation}
\begin{aligned}
\label{helampl}
&H_{\pm}(w) = f(w) \mp m_B m_{D^*} \sqrt{w^2-1}\,g(w),\,\,\,\,\,H_0(w) = \frac{\mathcal{F}_1(w)}{\sqrt{m_B^2+m_D^2-2m_Bm_Dw}},\\
& \hskip 2.5truecmH_t(w) = \frac{(m_B+m_{D^*}) \sqrt{m_B m_{D^*} (w^2-1)}}{\sqrt{m_B^2+m_{D^*}^2-2m_Bm_{D^*}w}}P_1(w).
\end{aligned}
\end{equation}
Note that $H_t(w)$ does not appear in the Eq.\,(\ref{finaldiff333BDst}) since its contribution is proportional to the lepton mass.

\begin{figure}[h!]
\begin{center}
\subfloat{\includegraphics[scale=0.5]{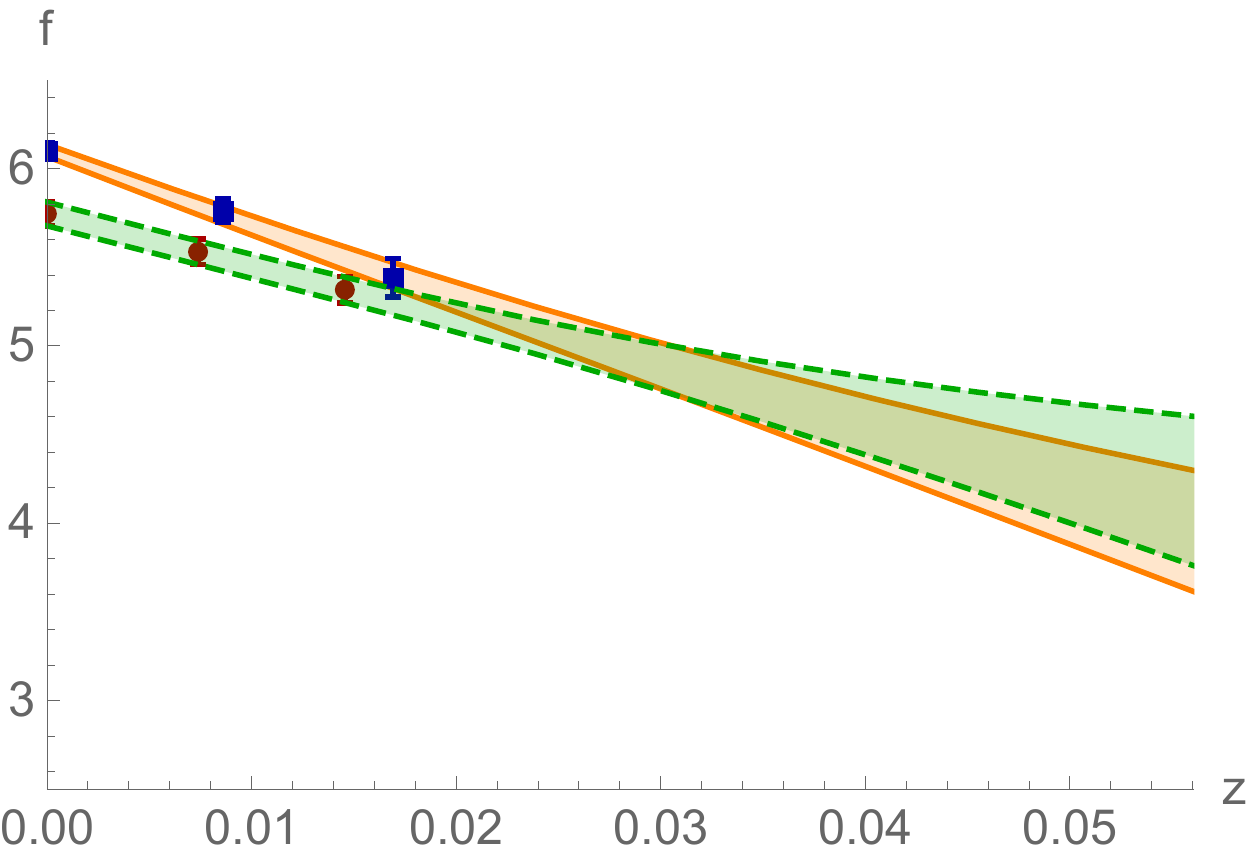}
\label{fig:A}}
\subfloat
{\includegraphics[scale=0.5]{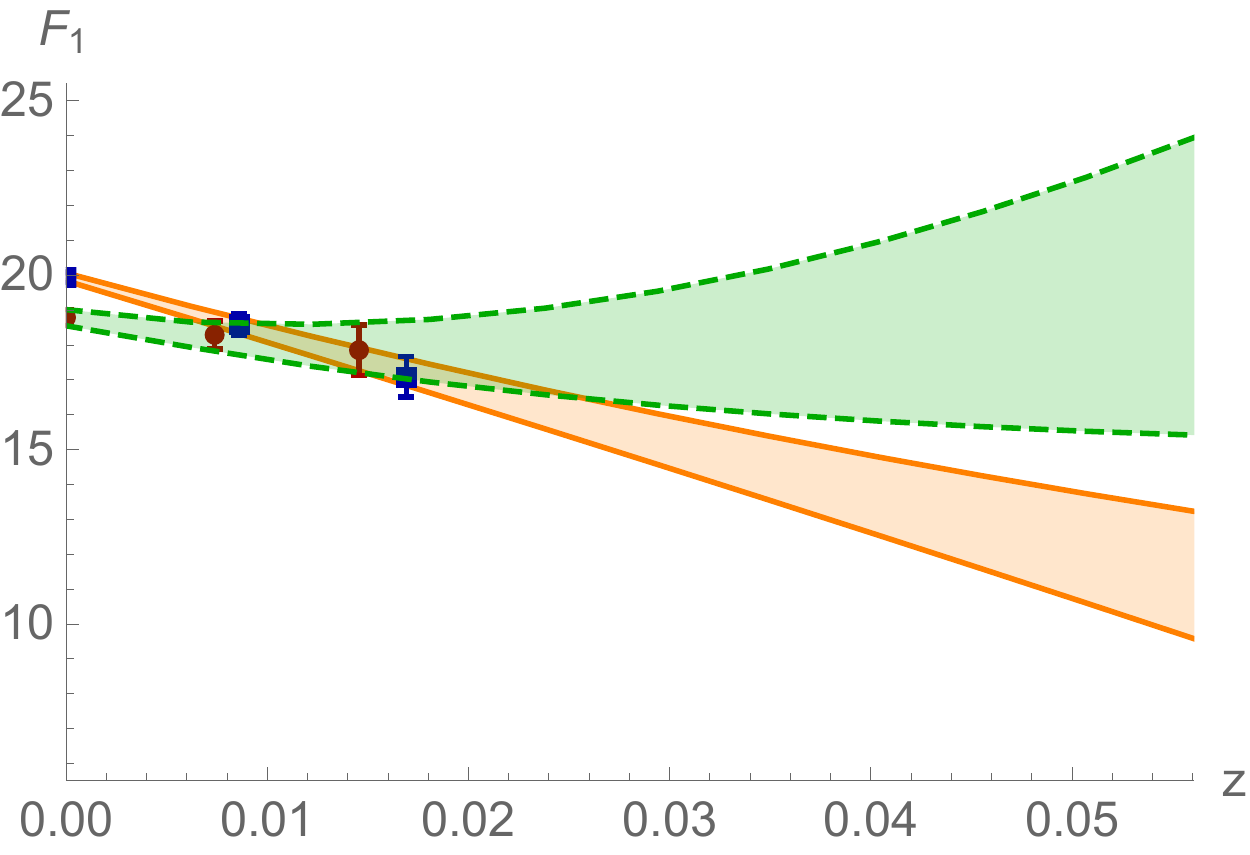}
\label{fig:B}}\\
\subfloat
{\includegraphics[scale=0.5]{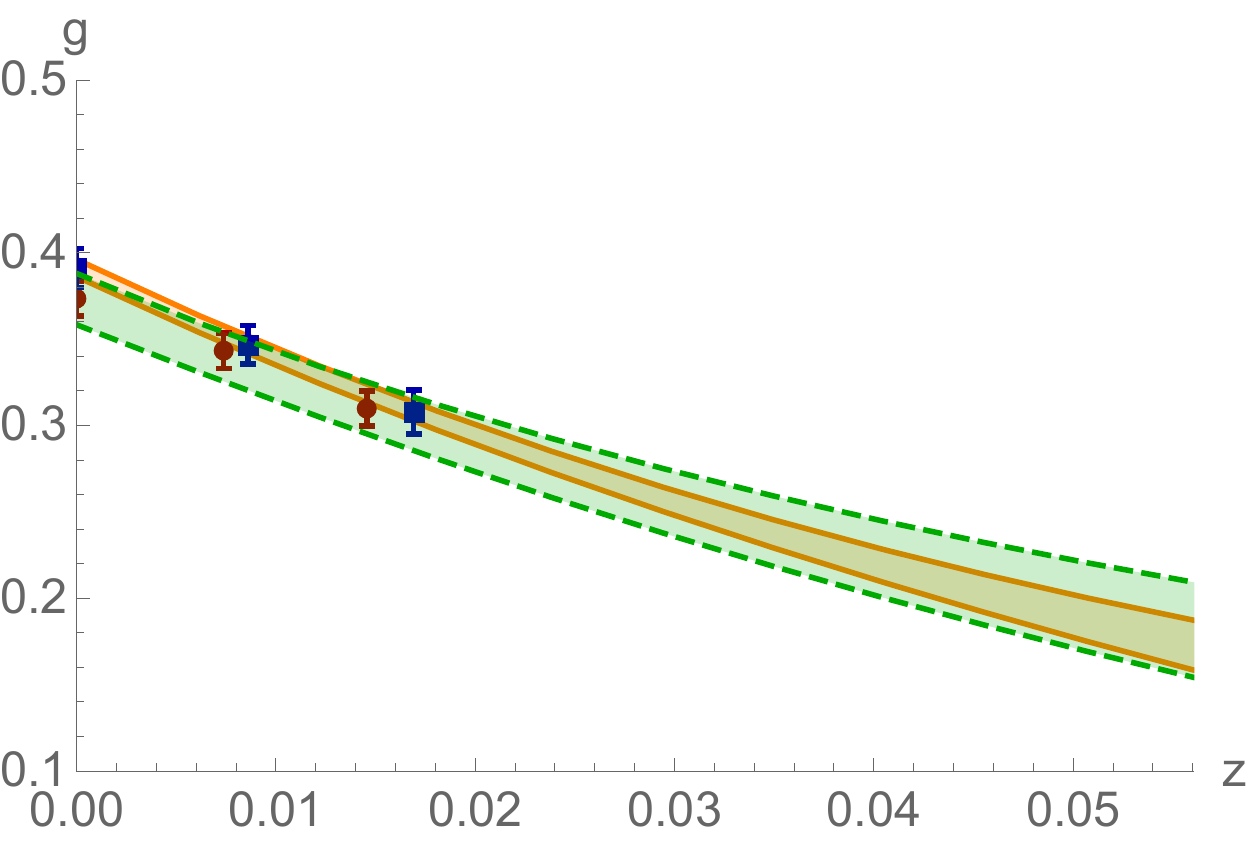}
\label{fig:C}}
\subfloat
{\includegraphics[scale=0.5]{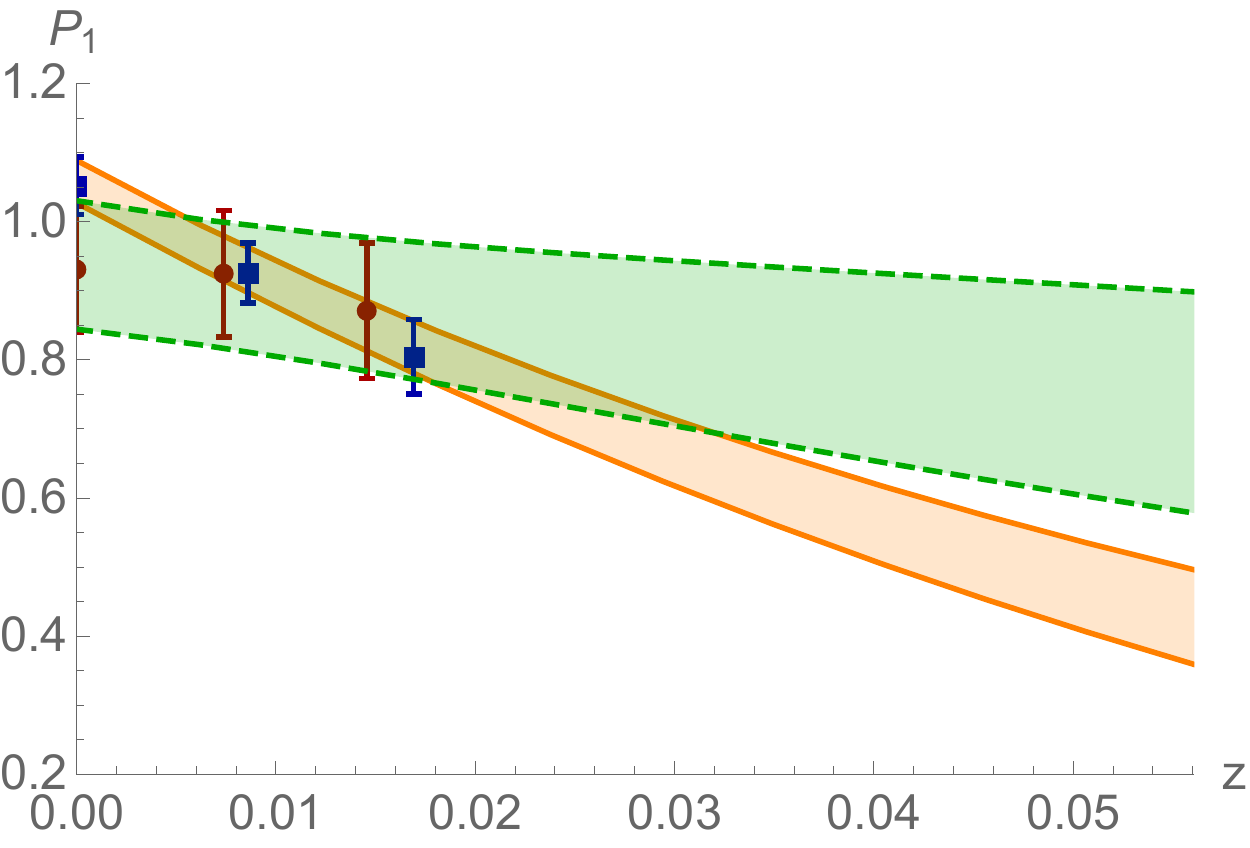}
\label{fig:D}}
\caption{\it\small The bands of the four FFs entering $B \to D^* \ell \nu$ decays, computed through the DM method. The solid orange (dashed green) bands are the results adopting the MILC (JLQCD) results, which are represented by blue squares (red points).
\hspace*{\fill}}
\label{FFMM}
\end{center}
\end{figure}

We have four FFs, $i.e.$ $f(w),g(w),\mathcal{F}_1(w),P_1(w)$, that we studied by using the Dispersive Matrix (DM) method, presented in \cite{DiCarlo:2021dzg}. The DM method allows us to study the FFs in a novel, non-perturbative and model-independent way, since, starting from the available LQCD computations of the FFs at low recoil, we can extrapolate their behaviour in the high-$w$ region. To this end, we do \textbf{not} assume any functional dependence of the FFs on the recoil and we use only non-perturbative quantities, including the susceptibilities (which have been computed for the first time on the lattice in \cite{Martinelli:2021frl} for $b \to c$ quark transitions). Furthermore, the resulting bands of the FFs will be independent of the experimental measurements of the differential decay widths, that will be then used only to obtain a new estimate of $\vert V_{cb} \vert$. Thus, by using the preliminary LQCD computations of the FFs by the FNAL/MILC \cite{Aviles-Casco:2019zop} and the JLQCD \cite{Kaneko:2019vkx} Collaborations, we have obtained the bands in Fig.\,\ref{FFMM} as a function of the conformal parameter
\begin{equation}
\label{conf}
z=\frac{\sqrt{w+1}-\sqrt{2}}{\sqrt{w+1}+\sqrt{2}}.
\end{equation}

\section{How to compute theoretical estimates of $\vert V_{cb} \vert$ and $R(D^*)$ through unitarity}

The Belle Collaboration \cite{Abdesselam:2017kjf, Waheed:2018djm} has published two differente sets of measurements of the differential decay widths $d\Gamma/dx$ ($x=w,\cos\theta_l,\cos\theta_v,\chi$) in the form of ten bins for each kinematical variable. Thus, we have computed the theoretical estimate of each one-dimensional differential decay width (starting from the complete expression in the Eq.\,(\ref{finaldiff333BDst}) and using the unitarity bands of the FFs in Fig.\,\ref{FFMM}) and we have compared them to the experimental data in order to obtain bin-per-bin estimates of $\vert V_{cb} \vert$. We have developed a separate analysis for each set of LQCD input data. 

\begin{figure}[h!]
\begin{center}
\subfloat{\includegraphics[scale=0.5]{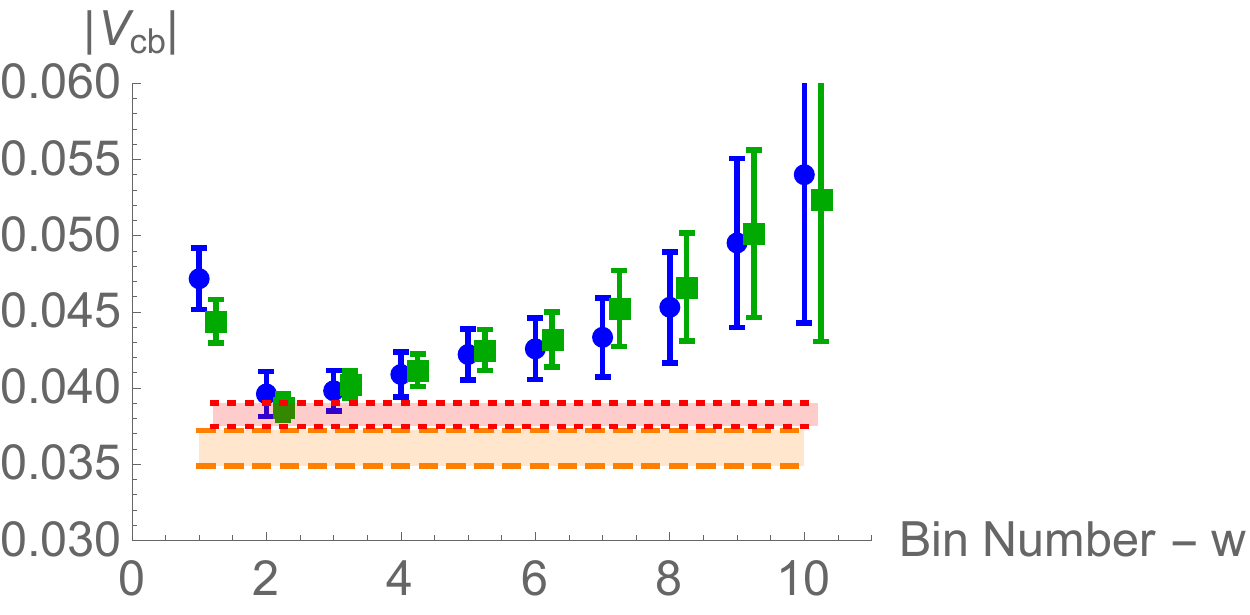}
\label{fig:A}}
\subfloat
{\includegraphics[scale=0.5]{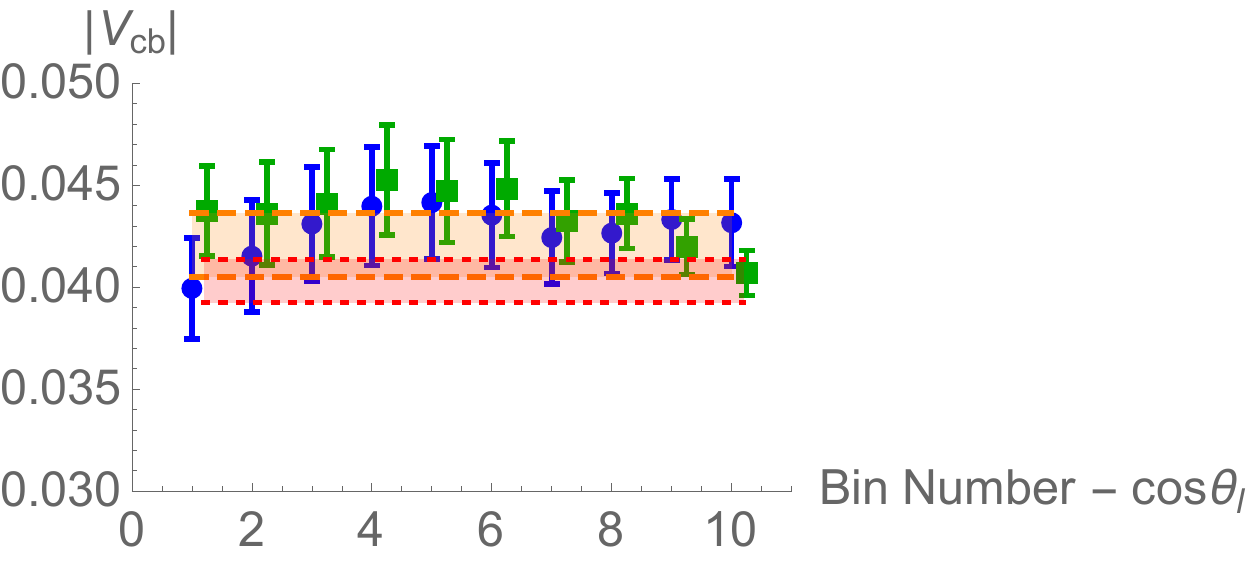}
\label{fig:B}}\\
\subfloat
{\includegraphics[scale=0.5]{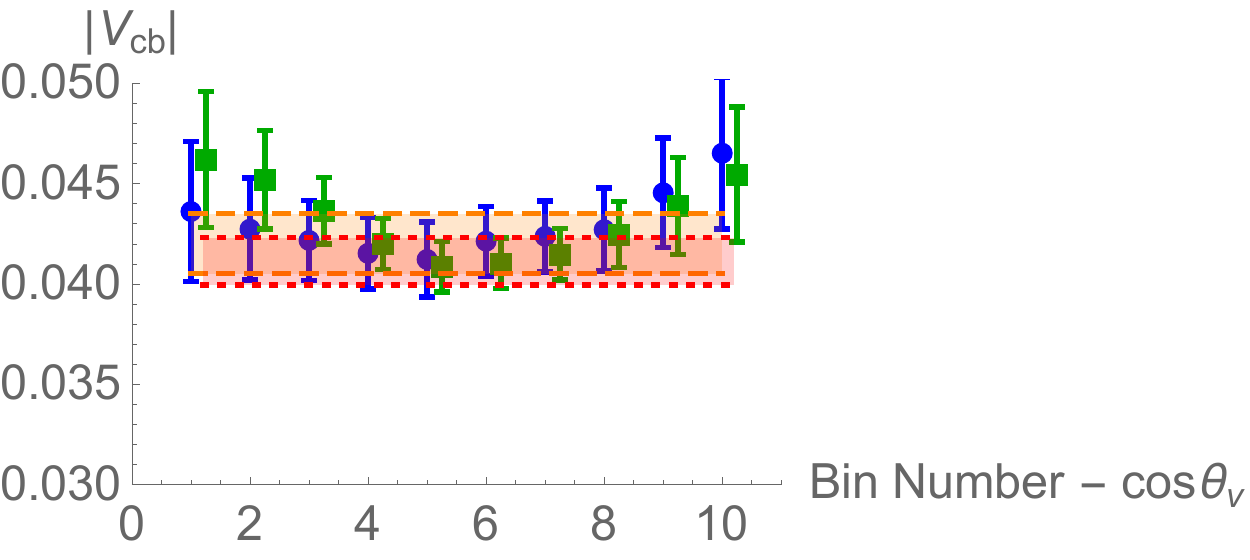}
\label{fig:C}}
\subfloat
{\includegraphics[scale=0.5]{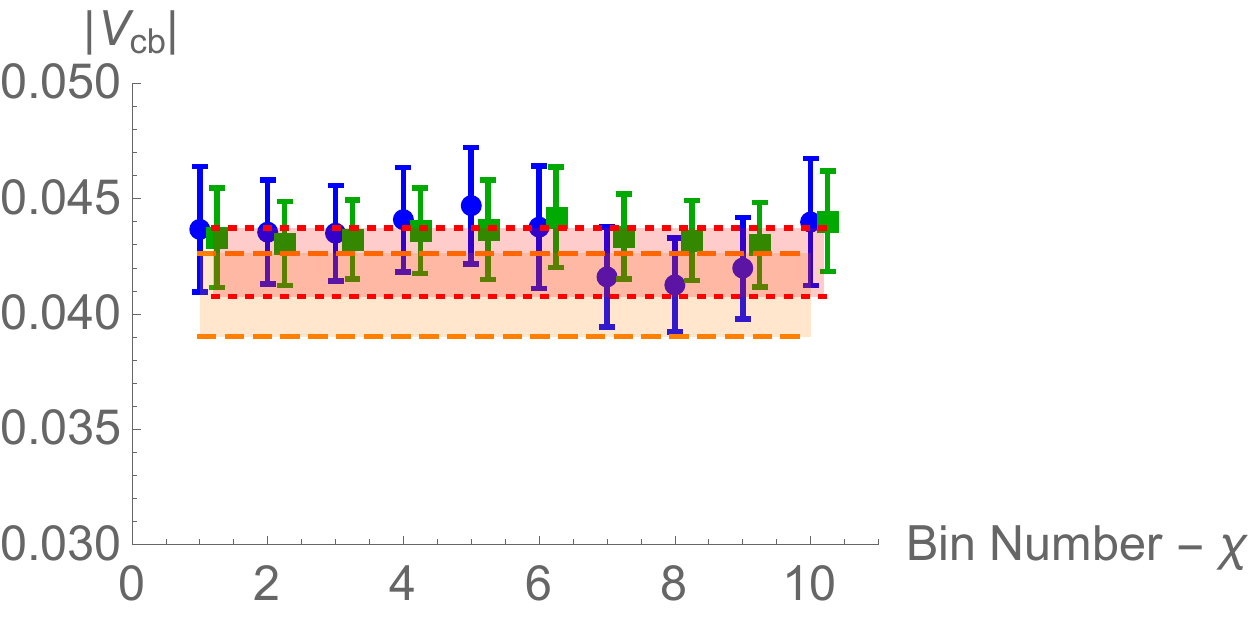}
\label{fig:D}}
\caption{\textit{Bin-per-bin values of $\vert V_{cb} \vert$ for each kinematical variable and for each Belle experiment, resulting from preliminary FNAL/MILC input data. The blue points correspond to the first Belle measurements \cite{Abdesselam:2017kjf}, while the green squares to the second one \cite{Waheed:2018djm}.  Finally, the dashed orange (dotted red) bands are the weighted mean values (\ref{muVcbfinal}) for each variable, starting from the blue points (green squares).}
\hspace*{\fill} \small}
\label{Vcb111}
\end{center}
\end{figure}

Let us examine for instance the results by adopting the FNAL/MILC preliminary LQCD inputs. They are shown in Fig.\,\ref{Vcb111}, where the blue points come from the first set of Belle measurements \cite{Abdesselam:2017kjf} while the green squares from the second one \cite{Waheed:2018djm}. For each kinematical variable and for each experiment, we compute the average of these bin-per-bin estimates as
\begin{equation}
\label{muVcbfinal}
\vert V_{cb} \vert = \frac{\sum_{i,j=1}^{10} (\mathbf{C}^{-1})_{ij} \vert V_{cb} \vert_j}{\sum_{i,j=1}^{10} (\mathbf{C}^{-1})_{ij}},\,\,\,\,\,\,\,\,\,\,\,\,\sigma^2_{\vert V_{cb} \vert} = \frac{1}{\sum_{i,j=1}^{10} (\mathbf{C}^{-1})_{ij}},
\end{equation}
where $\mathbf{C}$ is the covariance matrix and $\vert V_{cb} \vert_i$ are the single $\vert V_{cb} \vert$ estimates. This procedure generates the dashed orange (dotted red) bands in Fig.\,\ref{Vcb111} for the first (second) set of Belle measurements. As graphically clear, in the $w$ case there is an evident underestimate of the weighted mean value. 

\begin{figure}[h!]
\begin{center}
\subfloat{\includegraphics[scale=0.5]{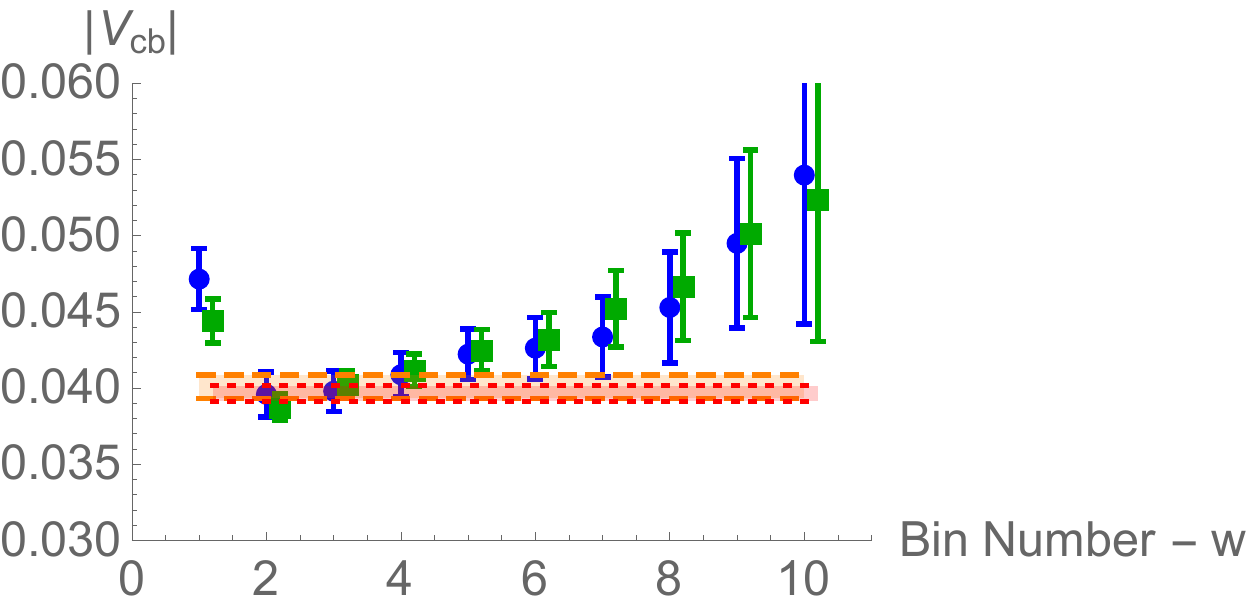}
\label{fig:A}}
\subfloat
{\includegraphics[scale=0.5]{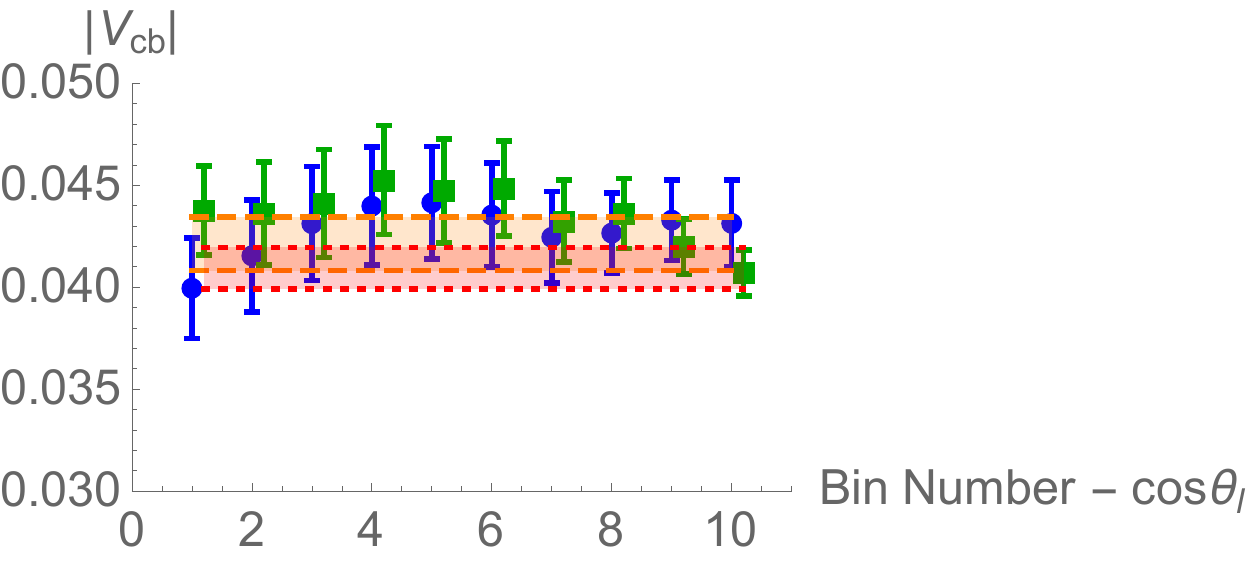}
\label{fig:B}}\\
\subfloat
{\includegraphics[scale=0.5]{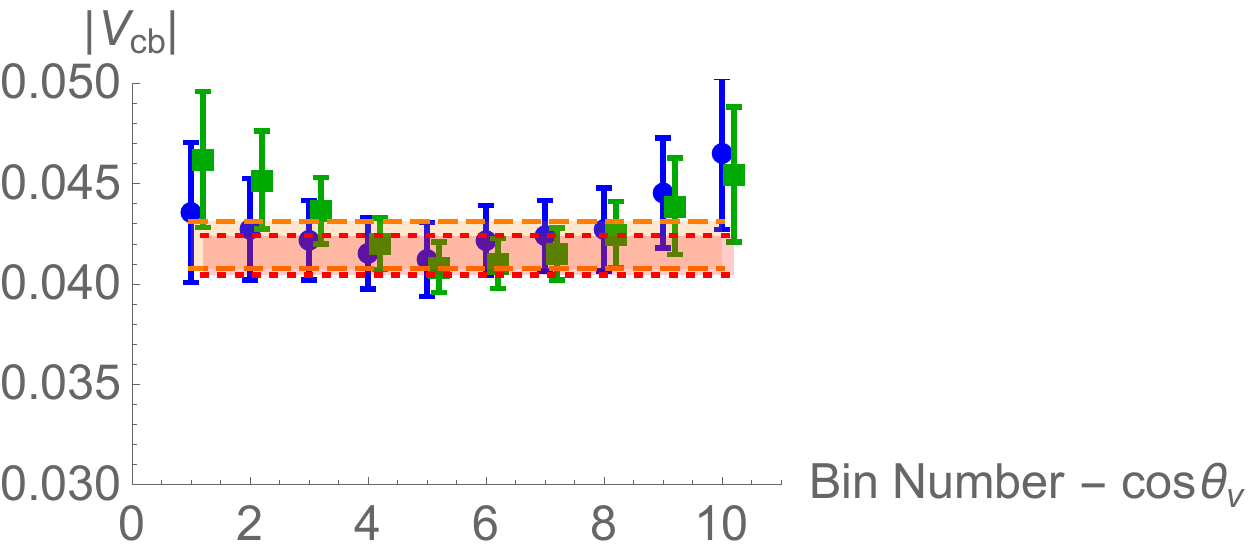}
\label{fig:C}}
\subfloat
{\includegraphics[scale=0.5]{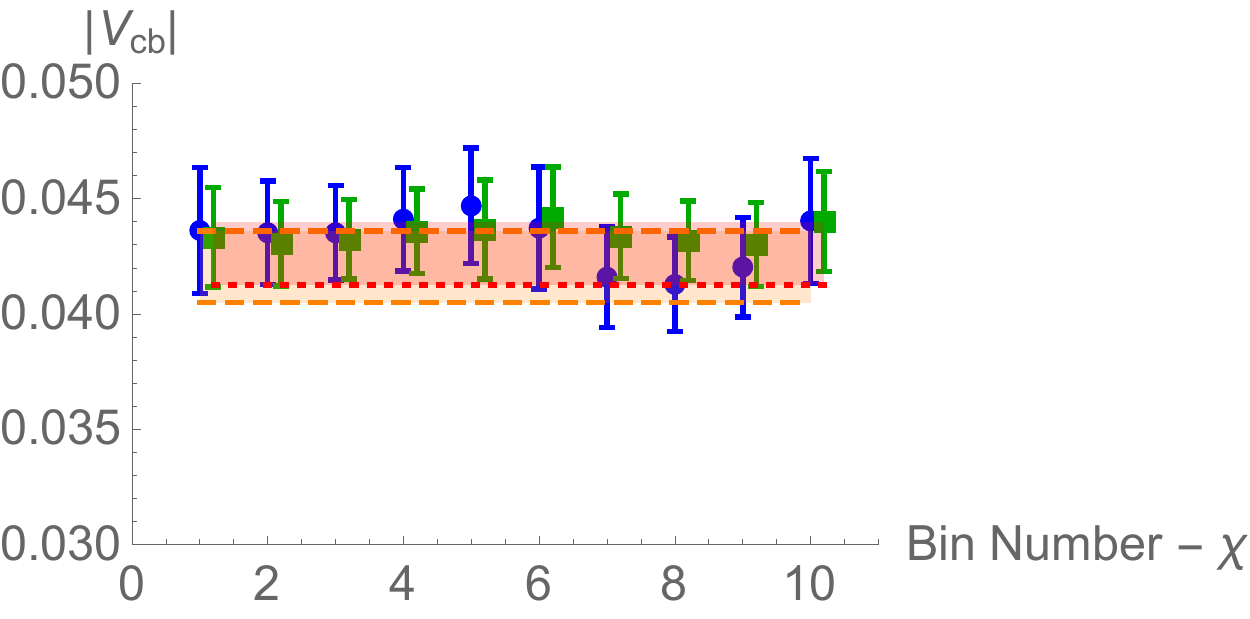}
\label{fig:D}}
\caption{\textit{Bin-per-bin values of $\vert V_{cb} \vert$ for each kinematical variable and for each Belle experiment, resulting from preliminary FNAL/MILC input data, after the modification of the experimental covariances. The colour code is the same one of Fig.\,\ref{Vcb111}.}
\hspace*{\fill} \small}
\label{Vcb2}
\end{center}
\end{figure}

To solve this problem, we consider the \emph{relative} differential decay widths, given by the ratios $(d\Gamma/dx)/\Gamma$, for each bin by using the measurements  \cite{Martinelli:2021onb}. We expect that the calibration errors in the experimental data will be strongly reduced in the ratios \cite{DAgostini:1993arp}. We compute a new correlation matrix $\rho$  through the generation of the bootstraps for $(d\Gamma/dx)/\Gamma$ and then a new covariance matrix $C_{new}$ of the measurements through the original experimental uncertainties $\sigma_{exp}$ associated to the data, $i.e.$ $C_{new,ij}=\rho_{ij} \sigma_{exp,i}\sigma_{exp,j}$.
We repeat the whole procedure for the extraction of $\vert V_{cb} \vert$ and compute the new weigthed mean values of $\vert V_{cb} \vert$ through the Eq.\,(\ref{muVcbfinal}). In Fig.\,\ref{Vcb2} we show the results. In this case, although differences remain, there is no systematic underestimate of the mean of $\vert V_{cb} \vert$. To obtain a final estimate of the CKM matrix element, we finally combine the resulting eight mean values as
\begin{eqnarray}
\label{sigma28}
\mu_{x} = \frac{1}{N} \sum_{k=1}^N x_k,\,\,\,\,\,\,\,\,\,\,\,\,\,\,\,\,\,\,\, \sigma^2_{x} = \frac{1}{N} \sum_{k=1}^N \sigma_k^2 + \frac{1}{N} \sum_{k=1}^N(x_k-\mu_{x})^2,
\end{eqnarray}
thus obtaining
\begin{equation*}
\vert V_{cb} \vert \times 10^{3} = 41.4 \pm 1.5
\end{equation*}
as the $\vert V_{cb} \vert$ estimate resulting from our study of the FNAL/MILC preliminary LQCD data.

The DM method allows us to obtain fully-theoretical expectation values of other quantities relevant for phenomenology, $i.e.$ the anomaly $R(D^*)$,
the $\tau$-polarization $P_{\tau}$ and finally the $D^*$ longitudinal polarization $F_L$. To be more specific, we generate some bootstrap events of the FFs and compute the corresponding values of $R(D^*)$, $P_{\tau}$ and $F_L$. The final means and the final uncertainties will come from a Gaussian fit of the histograms of these events. The results of our study of the FNAL/MILC preliminary LQCD data are
\begin{eqnarray*}
R(D^*) = 0.272 \pm 0.010,\,\,\,\,\,\,\,\,\,\,P_{\tau} = -0.52 \pm 0.01,\,\,\,\,\,\,\,\,\,\,F_L = 0.43 \pm 0.02.
\end{eqnarray*}

\section{Numerical results for $\vert V_{cb} \vert$ and $R(D^*)$ using the \emph{final} results by FNAL/MILC}

On May 2021, the FNAL/MILC Collaborations have published the final results of their LQCD computations of the FFs. For this reason, we have repeated the analysis described in the previous Section using the new data \cite{FermilabLattice:2021cdg}. Our final estimate for $\vert V_{cb} \vert$ then reads
\begin{equation}
\label{VcbD*}
\vert V_{cb} \vert \times 10^{3} = 41.3 \pm 1.7
\end{equation}
after the modification of the experimental covariances. Moreover, we obtain 
\begin{eqnarray}
\label{RD*}
R(D^*) = 0.269 \pm 0.008,\,\,\,\,\,\,\,\,\,\,P_{\tau} = -0.52 \pm 0.01,\,\,\,\,\,\,\,\,\,\,F_L = 0.42 \pm 0.01,
\end{eqnarray}
as final estimates of the anomaly and of the polarization observables \cite{Martinelli:2021myh}.

\section{Estimates of $\vert V_{cb} \vert$ and $R(D)$ from unitarity}

\begin{figure}[h!]
 \centering
\includegraphics[width=0.8\textwidth]{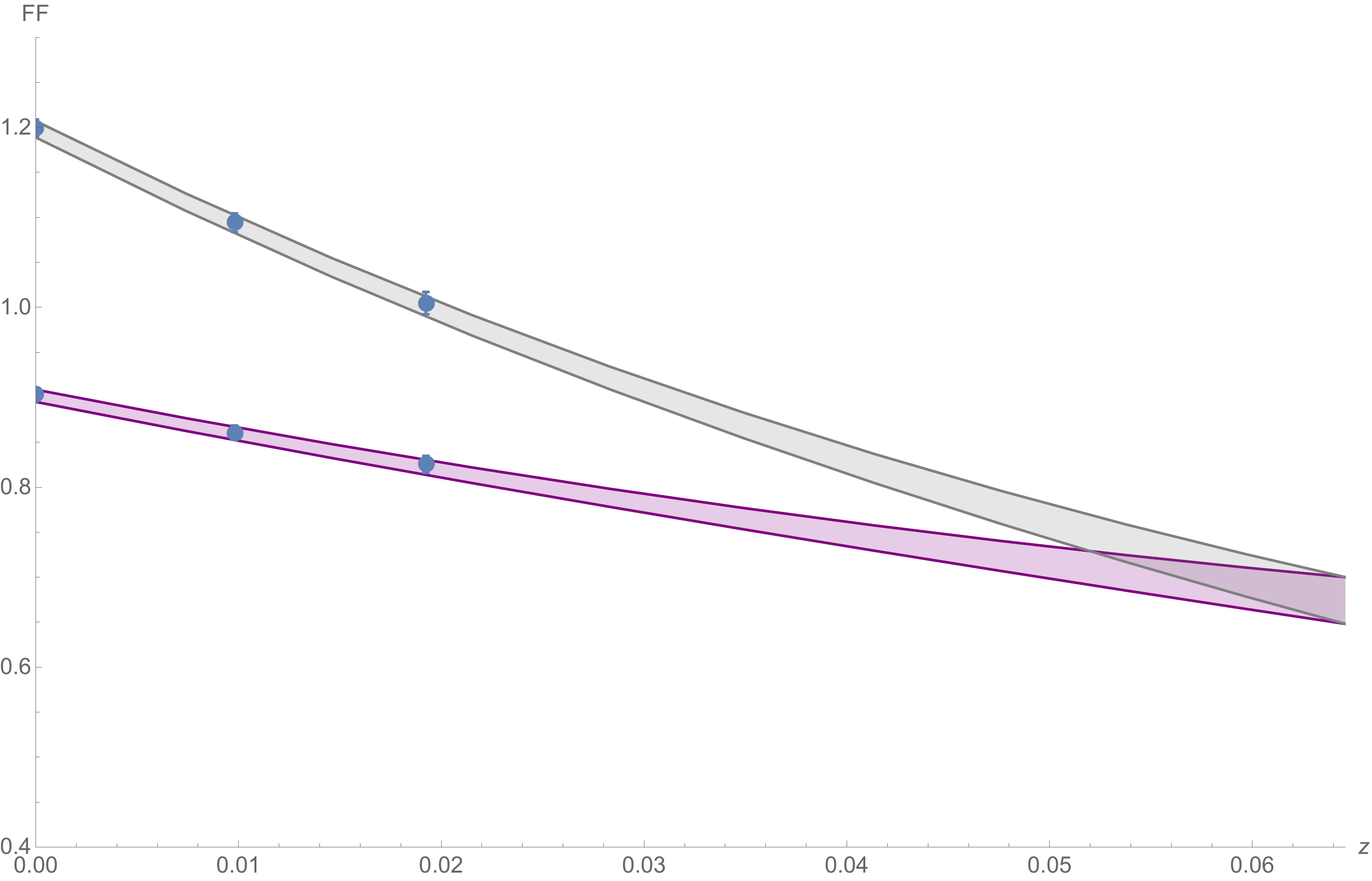}
 \centering
\caption{\textit{The bands of the FFs entering $B \to D\ell\nu$ decays computed through the DM method. The colour code is (lower) violet band for $f_0(z)$, (higher) gray one for $f_+(z)$. The blue points are the results of the FNAL/MILC computations.}
\hspace*{\fill} \small}
\label{FFsBD}
\end{figure}

The study of the semileptonic $B \to D$ decays is much simpler with respect to the $B \to D^*$ one, since the produced meson is a pseudoscalar one. In particular, we have to deal only with the momentum transfer as a kinematical variable, so that we can express the differential decay width as
\begin{equation}
\begin{aligned}
\label{finaldiff333}
&\frac{d\Gamma}{dq^2}=\frac{G_F^2 \vert V_{cb} \vert^2 \eta_{EW}^2}{24\pi^3} \left(1-\frac{m_{\ell}^2}{q^2}\right)^2\\
&\hskip 0.65truecm\times \left[\vert \vec{p}_{D}\vert^3 \left(1+\frac{m_{\ell}^2}{2q^2}\right) \vert f^+(q^2) \vert^2 + m_{B}^2 \vert \vec{p}_{D} \vert \left( 1-\frac{m_{D}^2}{m_{B}^2}\right)^2 \frac{3m_{\ell}^2}{8q^2} \vert f^0(q^2) \vert^2\right],
\end{aligned}
\end{equation}
where the two FFs $f_{+,0}(q^2)$ are related through the expression $f^{+}(0) = f^{0}(0)$. Starting from the available LQCD computations of these FFs by the FNAL/MILC Collaborations  \cite{Bailey_2015} and from our non-perturbative susceptibilities \cite{Martinelli:2021frl}, the DM method allows us to describe the FFs in the whole kinematical range. Our results are shown in Fig.\,\ref{FFsBD}.

\begin{figure}[h!]
 \centering
\includegraphics[width=0.6\textwidth]{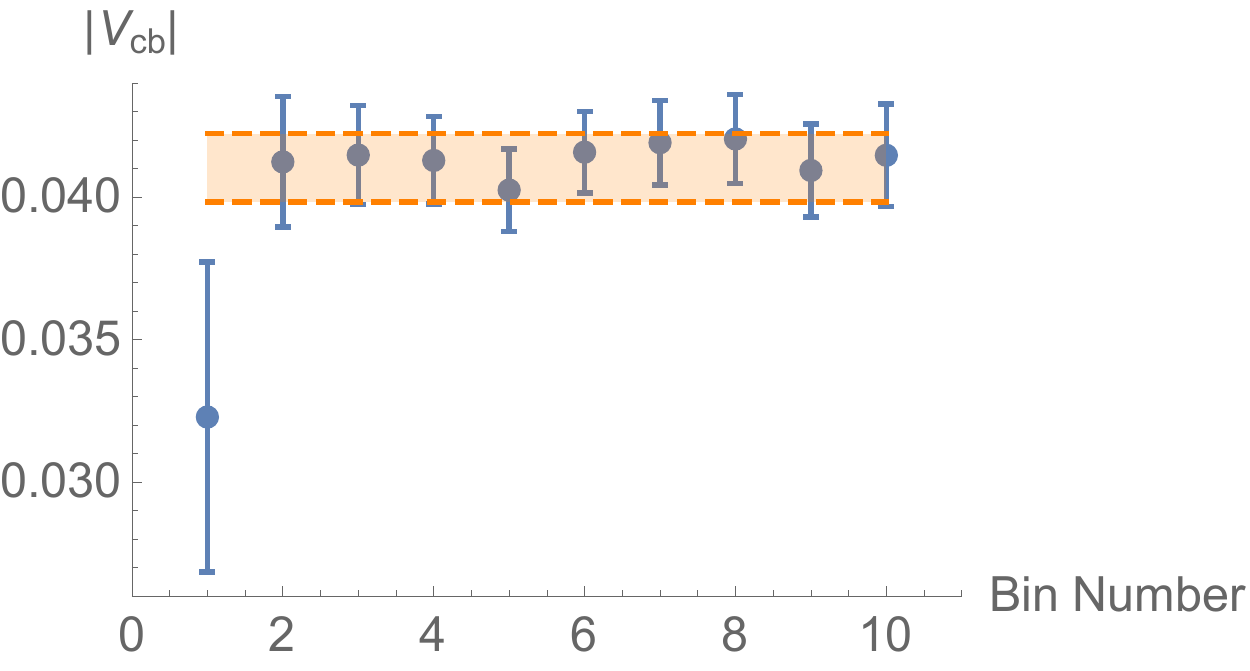}
 \centering
\caption{\textit{Values of the CKM matrix element $\vert V_{cb} \vert$ resulting from the experimental data for semileptonic $B \to D$ decays. The orange band represents the result of the weighted mean (\ref{VcbBD}).}
\hspace*{\fill} \small}
\label{VcbBDfig}
\end{figure}

In order to obtain an updated value of $\vert V_{cb} \vert$ from this channel, we put together our description of the lattice FFs and the measurements of the differential decay width $d\Gamma/dw$. For the latter case, we use the results published by the Belle Collaboration \cite{Glattauer_2016}, for which they also give the correlation matrix of the systematic errors. We re-express the Eq.\,(\ref{finaldiff333}) as
\begin{equation}
\label{VcbFINAL}
\vert V_{cb} \vert = \sqrt{\frac{d\Gamma}{dq^2}\vert_{exp} \times \frac{24 \pi^3}{G_F^2 \eta_{EW}^2 \vert \vec{p}_{D} \vert^3 \vert f^+(q^2) \vert^2_{th}}},
\end{equation}
where we are neglecting the lepton mass for simplicity. By using this expression, we compute $\vert V_{cb} \vert$ for each recoil bin through a bootstrap analysis, as shown in Fig.\,\ref{VcbBDfig}, and Eq.\,(\ref{muVcbfinal}) allows us to obtain a final estimate of $\vert V_{cb} \vert$, namely 
\begin{equation}
\label{VcbBD}
\vert V_{cb} \vert = (41.0 \pm 1.2) \cdot 10^{-3},
\end{equation}
which is represented as an orange band in the same Figure.

Finally, we compute a new theoretical estimate of $R(D)$ by using the same strategy for $R(D^*)$, $P_{\tau}$ and $F_L$. Analogously to those cases, we generate bootstrap events of the FFs $f_{+,0}$ in the whole kinematical range and thus we determine the corresponding values of $R(D)$, namely
\begin{equation}
\label{RD}
R(D) = 0.289(8).
\end{equation}

\section{Conclusions}

We have studied the semileptonic $B \to D^{(*)}\ell\nu$ decays through the Dispersive Matrix description of the Form Factors. Our main results are given in Eqs.\,(\ref{VcbD*}) and (\ref{VcbBD}) for the extraction of $\vert V_{cb} \vert$ from the $B \to D^*$ and the $B \to D$ decays, respectively. We have for the first time an indication of a sizable reduction of the  tension between the inclusive and the exclusive determinations of $\vert V_{cb} \vert$ in both the channels. Furthermore, our results for the anomalies are given in Eqs.\,(\ref{RD*}) and (\ref{RD}). With respect to other phenomenological studies in literature, the uncertainties on the ratios $R(D^{(*)})$ are larger since the DM method is completely model-independent and only LQCD data are used to describe the shape of the FFs. The difference between these theoretical expectations and the corresponding experimental world averages is reduced to $\sim1.6\sigma$ level.


\bibliographystyle{JHEP}
\bibliography{notes_biblio}

\end{document}